\documentclass{article}

\usepackage{jheppub}
\usepackage[utf8]{inputenc}
\usepackage{multirow}
\usepackage[toc,page]{appendix}

\usepackage{amsmath}
\usepackage{amsfonts}
\usepackage{amssymb}
\usepackage{graphicx}
\usepackage{ulem} 
\usepackage{slashed}
\usepackage{doi}
\usepackage{hyperref}
\usepackage{float}
\usepackage{csquotes}

\usepackage{graphicx}
\usepackage{caption}
\usepackage{subcaption}
\usepackage{booktabs}
\usepackage{cleveref}

\usepackage[ruled,vlined]{algorithm2e}
\usepackage[most]{tcolorbox}

\usepackage{placeins}
\usepackage{ulem}
\usepackage{xcolor}

\makeatletter
\gdef\@fpheader{\vline}
\makeatother

\author[a,b]{Sascha Caron}
\emailAdd{scaron@nikhef.nl}
\author[a,b]{Polina Moskvitina}
\emailAdd{p.moskvitina@nikhef.nl}
\author[c]{Roberto Ruiz de Austri}
\emailAdd{rruiz@ific.uv.es}
\author[a,b]{Eugene Shalugin}
\emailAdd{eugene.shalugin@ru.nl}

\affiliation[a]{Nikhef, Dutch National Institute for Subatomic Physics, Science Park 105, 1098 XG Amsterdam, The Netherlands}
\affiliation[b]{High Energy Physics, Radboud University Nijmegen, Heyendaalseweg 135, 6525 AJ Nijmegen, The Netherlands}
\affiliation[c]{Instituto de Física Corpuscular, IFIC-UV/CSIC, Catedrático José Beltrán 2, E-46980 Paterna, Spain}

\title{Hits to Higgs: Hit-Level Higgs Classification from Raw LHC Detector Data Using Higgsformer}

\abstract{
We present Higgsformer, a transformer-based architecture that classifies Higgs events at the Large Hadron Collider directly from raw inner tracker hits, bypassing the traditional reconstruction chain of intermediate physics objects.
As a benchmark, we focus on distinguishing $t\bar{t}H$ from $t\bar{t}$ events with $H \to b\bar{b}$, a particularly challenging task due to their similar final state topologies. Our pipeline begins with event generation in \texttt{Pythia8}, fast simulation with ACTS/Fatras, and classification directly from raw detector hits.

We show for the first time that a transformer model originally developed for inner tracker hit-to-track assignment can be retrained to classify Higgs signal events directly from raw hits. For comparison, we reconstruct the same events with \texttt{Delphes} and train a Particle Transformer as an object-based classifier. We evaluate both approaches under varying dataset sizes and pileup levels. Despite relying exclusively on inner tracker hits, our large Higgsformer achieves an AUC of $0.855$, matching the performance of the traditional reconstruction pipeline at a $b$-tagging efficiency of $\approx 40\%$  under the same detector constraints.
}

\begin{document}

\maketitle


\section{Introduction}

\begin{figure}[h]
    \centering
    \includegraphics[scale=0.45]{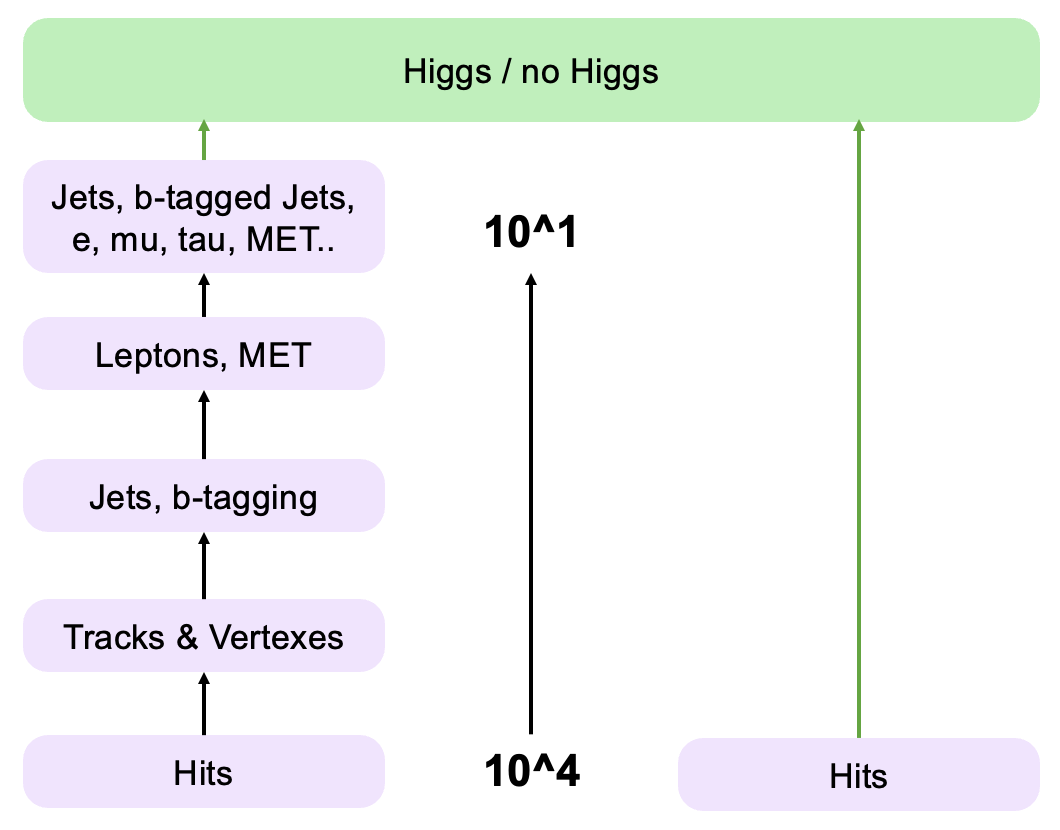}
    \caption{Reconstruction pipeline overview, compared to learning from raw detector data. Classical reconstruction pipeline compresses $10^4$ raw detector hits per event to $10^1$ reconstructed objects.}
    \label{fig:compression}
\end{figure}

The Large Hadron Collider (LHC) at CERN produces proton-proton collisions at an unprecedented scale, up to $40$ million events per second, generating hundreds of terabytes of online data and tens of petabytes of stored offline data annually. Traditionally, this massive volume is processed through deeply structured reconstruction pipelines. These include particle trajectory fitting, calorimeter clustering, and jet and object identification, which ultimately reduce the raw detector output into a compact set of high-level observables used in physics analyses as shown in~\Cref{fig:compression}. While effective, this approach imposes strong inductive biases and may discard low-level information that could be critical for certain tasks.

This raises a natural question: \textit{Can modern machine learning models learn directly from the raw detector data, bypassing intermediate reconstruction stages and high-level features altogether?} In this work, we address this question by directly comparing two pipelines: one based on an end-to-end classifier operating on raw digitised detector hits obtained with ACTS/Fatras, and another using a baseline classifier trained on high-level reconstructed objects obtained from a Delphes-simulated detector. Both pipelines share identical physics event generation to ensure a fair comparison.

The classification task we consider is the separation of $t\bar{t}$ from $t\bar{t}H$ events, with the Higgs boson forced to decay to a $b\bar{b}$ pair. 
Discriminating $t\bar{t}H$ events from the dominant $t\bar{t}$ background, especially in the $H \to b\bar{b}$ decay channel, is a challenging task in Higgs physics at the LHC~\cite{ATLAS:2024gth,CMS:2024fdo}.

This pair of processes provides a realistic and challenging benchmark: they share similar final state topologies, but differ subtly in object multiplicities and kinematics, particularly in the presence of additional $b$-jets from the Higgs decay. These subtle differences make it an ideal case study for assessing whether low-level hit information can rival or surpass traditional high-level approaches.


We build a complete end-to-end event simulation pipeline based on \texttt{Pythia8}~\cite{Sj_strand_2015} event generation, ACTS/Fatras~\cite{Ai:2021ghi} detector simulation, and a FlashAttention-accelerated~\cite{dao2023flashattention2} transformer model to classify events directly from hit patterns. 
A closely related model was previously developed by us as a \texttt{Trackformer}, used for assigning hits to tracks~\cite{Caron:2024cyo}. In this work, we adapt the same basic architecture for Higgs event classification and refer to it as the \texttt{Higgsformer}. It is a lightweight, set-based Transformer designed to operate directly on raw detector hits. In addition, we demonstrate the capabilities of a larger model.


For comparison, we construct a parallel baseline in which the same generated events are processed through \texttt{Delphes}~\cite{deFavereau:2013fsa} to reconstruct track-jets, their $b$-tag information, and track-based missing transverse momentum.
On this object-level representation, we benchmark against 
the Particle Transformer (ParT) \cite{Qu:2022mxj}. This setup allows us to highlight the novelty of hit-level end-to-end learning with the Higgsformer, while still enabling a comparison to established object-level approaches. 
We then evaluate both pipelines under varying detector conditions, including datasets with and without additional proton--proton interactions (pileup), to assess robustness in realistic LHC-like environments. 

We find that the hit-level model achieves performance competitive with, and in some cases approaching, that of the reconstructed object baseline, despite bypassing the entire reconstruction chain. This result supports the potential of hit-level end-to-end learning in high energy physics.

We emphasise that this work is a simulation-based proof-of-concept and does not establish a complete deployment strategy on collision data. Hit-level classifiers can be more sensitive to residual data--simulation differences than approaches based on standard reconstructed objects with mature calibration/validation programs. A realistic application would therefore require integration into the experiment’s standard workflow, together with robustness studies, calibrations and data-driven tests in control samples and regions. This is beyond the scope of the present paper and left for future work.

The remainder of this paper is organised as follows. Section~\ref{sec:work} reviews related work. Section~\ref{sec:dataset} describes the event generation procedure and the two parallel datasets: the hit-level dataset obtained with ACTS/Fatras and the reconstructed object dataset obtained with \texttt{Delphes}. Section~\ref{sec:ml-models} introduces the machine learning architectures employed in this study, with particular emphasis on our novel Higgsformer model. Section~\ref{sec:experiments} presents the experimental setup and comparative results, covering performance as a function of dataset size, pileup conditions, and learned feature exploration. Finally, Section~\ref{sec:conclusions} summarises our conclusions and outlines directions for future work.

\section{Related Work} 
\label{sec:work}

The classification of $t\bar{t}H$ events particularly in the $H \rightarrow b\bar{b}$ decay channel has been the focus of various machine learning (ML) efforts in high energy physics. Prior work has explored both traditional algorithms and neural approaches, typically relying on handcrafted features derived from reconstructed-level data~\cite{ATLAS:2018mme,CMS:2018uxb}. 

V{\aa}ge~\cite{Vage2018} examined the performance of Boosted Decision Trees (BDTs) and Neural Networks (NNs) in the context of $t\bar{t}H(b\bar{b})$ classification at ATLAS. Using standard high-level features such as jet multiplicities, $b$-tag scores, and invariant masses, the study found comparable results between BDTs and fully connected NNs, with both achieving an area under the receiver operating characteristic (ROC) curve, or AUC, in the range of $75–77\%$. 
However, it also highlighted that neural networks are more sensitive to class imbalance and training biases, especially in low-statistics regions.

Similarly, Santos et al.~\cite{Santos:2016kno} conducted a broad benchmarking study of several ML classifiers, ranging from K-Nearest Neighbors and Naive Bayes to XGBoost and neural network variants, on simulated $t\bar{t}H$ events with $H \rightarrow b\bar{b}$. Their best models (XGBoost and NeuroBGD) achieved AUCs around $80.0–80.2\%$ and $F$-scores around $74\%$, but all required pre-engineered input variables and physics-specific observables.

While these studies demonstrated the power of ML in this classification task, they all operated on reconstructed objects such as jets, leptons, and missing transverse energy (MET). That paradigm implicitly assumes the availability of full event reconstruction, and carries potential biases from detector modelling, jet clustering algorithms, or $b$-tagging procedures.

A complementary line of work uses ML to perform particle- or object-level reconstruction starting from low-level detector information, providing a middle ground between conventional object-based pipelines and direct hit-level classification. Examples include learning-based tracking that reconstructs charged-particle trajectories from hits via learned hit association and track building~\cite{Ju:2021exatrkx}, ML-assisted calorimeter clustering/shower reconstruction in high-granularity calorimeters~\cite{Bhattacharya:2022hgcal}, and ML-based particle-flow reconstruction that predicts a set of particle candidates from tracks and calorimeter clusters~\cite{Pata:2021mlpf}. These approaches explicitly produce reconstructed particles/objects and can therefore interface naturally with established calibration and validation workflows.

In contrast, our work takes a radically different approach. We eliminate the reconstruction step entirely and operate directly on the raw detector hits using transformer-based models. Inspired by recent work on TrackML~\cite{Caron:2024cyo,Kiehn:2019tbl}, we hypothesize that deep architectures can learn to extract task-relevant structure directly from hit patterns given enough training data and realistic detector simulation. 
To the best of our knowledge, this is the first study exploring end-to-end event-level classification at the LHC experiment, in particular $t\bar{t}$ vs. $t\bar{t}H$ events, from hit-level data alone, without using any handcrafted features or reconstructed quantities.

\section{Data Preparation}
\label{sec:dataset}

\subsection{Event Generation}
\label{ref:event-gen}

We simulate signal and background processes using \texttt{Pythia8} integrated within the \texttt{ACTS} framework. 

For the signal, we consider $t\bar{t}H$ events where the Higgs boson is forced to decay into a $b\bar{b}$ pair; a dominant decay channel for $m_H \approx 125$\,GeV and of high relevance for experimental analyses. We include both gluon-gluon fusion and quark-antiquark annihilation production channels:
\[
gg \rightarrow t\bar{t}H,\quad q\bar{q} \rightarrow t\bar{t}H \, .
\]
Initial and final state radiation (ISR/FSR) are enabled to simulate realistic parton showering. We apply a parton-level transverse momentum cut of $\hat{p}_T > 100$\,GeV to suppress soft events. Hadron-level decays are fully enabled for both the Higgs and top quarks.

The background consists of standard $t\bar{t}$ production, including QCD-induced additional jet radiation. This class includes light-flavour jets but lacks the explicit $b\bar{b}$ pair from Higgs decay.

Each event is generated with Gaussian vertex smearing to reflect LHC beam spot properties:
\[
\sigma_{x,y} = 12.5~\mu\mathrm{m}, \quad \sigma_z = 55.5~\mathrm{mm}, \quad \sigma_t = 5~\mathrm{ns}
\]
A fixed random seed of $42$ is used for reproducibility. All datasets are produced in both \texttt{ROOT} and \texttt{CSV} formats, with $20\,000$ signal and $20\,000$ background events per configuration.


\subsection{Pileup, postprcessing and reconstruction pipeline} 
\begin{figure}[h]
    \centering
    \includegraphics[scale=0.31]{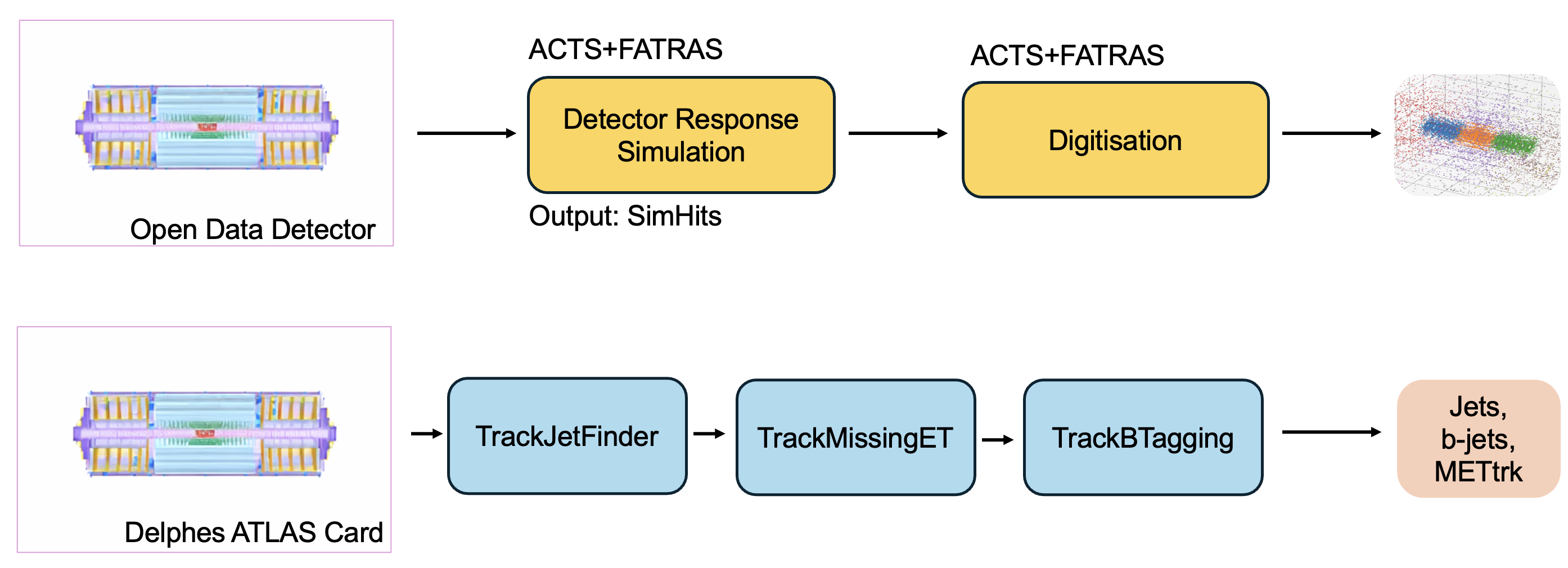}
    \caption{Pipeline for hit-level data generation with Open Data Detector and ACTS (up), and reconstructed objects-level data generation with Delphes (bottom).}
    \label{fig:data_pipelines}
\end{figure}

We generate two different datasets: a hit-level dataset and a reconstructed objects-level dataset. The generation pipeline is shown in~\Cref{fig:data_pipelines}.

We use the Fatras fast simulation engine provided by ACTS to simulate the interaction of particles with a tracking detector under a $2~T$ magnetic field. Particle propagation through the detector geometry includes effects such as multiple scattering, energy loss, and material interaction.
The detector geometry is defined using the ACTS \texttt{GenericDetector}, mirroring the configuration used in the TrackML challenge~\cite{Kiehn:2019tbl}. Fatras outputs truth-level hits (\texttt{SimHits}) for each detector surface where a particle intersects.
Following simulation, we perform digitisation to model sensor response. Digitisation is configured using ACTS JSON-based configuration files to emulate realistic pixel and strip detector behaviour. The result is a set of digitised hits stored in \texttt{CSV} format, suitable for downstream processing and machine learning input.


To create TrackML-like datasets, we first perform a coordinate transformation from local sensor coordinates to global $(x,y,z)$, then tag each hit with its volume, layer, and module indices. We also compute the transverse momentum, count the number of hits per track ($n_{\text{hits}}$), and remove zero-charge particles or malformed hits. Finally, inspired by TrackML, we assign per-hit weights in the same manner.

We generated datasets with pileup (PU) levels of $0$, $5$, and $20$ interactions per bunch crossing, each containing $40\,000$ events.\footnote{The full dataset generation pipeline is available at \href{https://github.com/EugeneShalli/hits-gen}{Hits-Gen}. The dataset for pileup 0 is available in Ref.~\cite{caron_2026_20056016}.} This enables the evaluation under varying detector occupancy and complexity.


For the object-level comparison we simulate detector effects with \texttt{Delphes}~3.5.0~\cite{deFavereau:2013fsa} using the default ATLAS card. 
We define an inner tracker-only baseline to better match the information content available to Higgsformer. In this tracker-only setup we restrict the inputs to track-based objects: track-jets reconstructed with the \texttt{TrackJetFinder} module and track-based missing transverse momentum from \texttt{TrackMissingET}; reconstructed leptons and photons are not included since they rely on calorimeter and/or muon-system information. Track-jet $b$-tagging follows the ATLAS-card implementation via the \texttt{TrackBTagging} module, which assigns a tag bit to each track-jet using a $p_T$-dependent efficiency parameterisation for $b$-jets (flavour index \texttt{\{5\}} in the card),
\begin{equation}
\epsilon_b(p_T)=f\,\tanh(0.003\,p_T)\left(\frac{30}{1+0.086\,p_T}\right),
\end{equation}
where $f$ is an overall prefactor that approximately sets the working point. We assess the sensitivity of the object-level baselines to the assumed tagging performance by varying only this $b$-jet working point, considering $f=80\%, 60\%, 40\%$.\footnote{This \texttt{Delphes} implementation is a simplified, fast-simulation parameterisation meant to emulate typical fixed-efficiency $b$-tagging operating points. In ATLAS analyses, $b$-tagging is performed with multivariate taggers and the corresponding efficiencies are measured in collision data and applied to simulation via data-to-simulation scale factors as functions of jet kinematics and event conditions; see e.g.\ Ref.~\cite{ATLAS:2019btag}. Track-jet $b$-tagging in boosted topologies (track-jets matched to large-$R$ calorimeter jets) is used and validated in ATLAS as well; see e.g.\ Ref.~\cite{ATLAS:2019boostedHbb}.}

The output consists of reconstructed high-level physics objects
\begin{itemize}
\item Track-jets with four-momenta $(p_T,\eta,\phi,E)$ and $b$-tag discriminants.
\item Track-based missing transverse momentum $\slashed{E}_T$ and its azimuthal angle.
\end{itemize}

Objects are ordered by decreasing $p_T$ and padded to a fixed multiplicity per category to enable batch processing in the ML pipeline. Missing entries are zero-padded, and a corresponding mask array is stored for each event.

\section{Machine Learning Models}
\label{sec:ml-models}
\begin{figure}[h]
    \centering
    \includegraphics[width=1\linewidth]{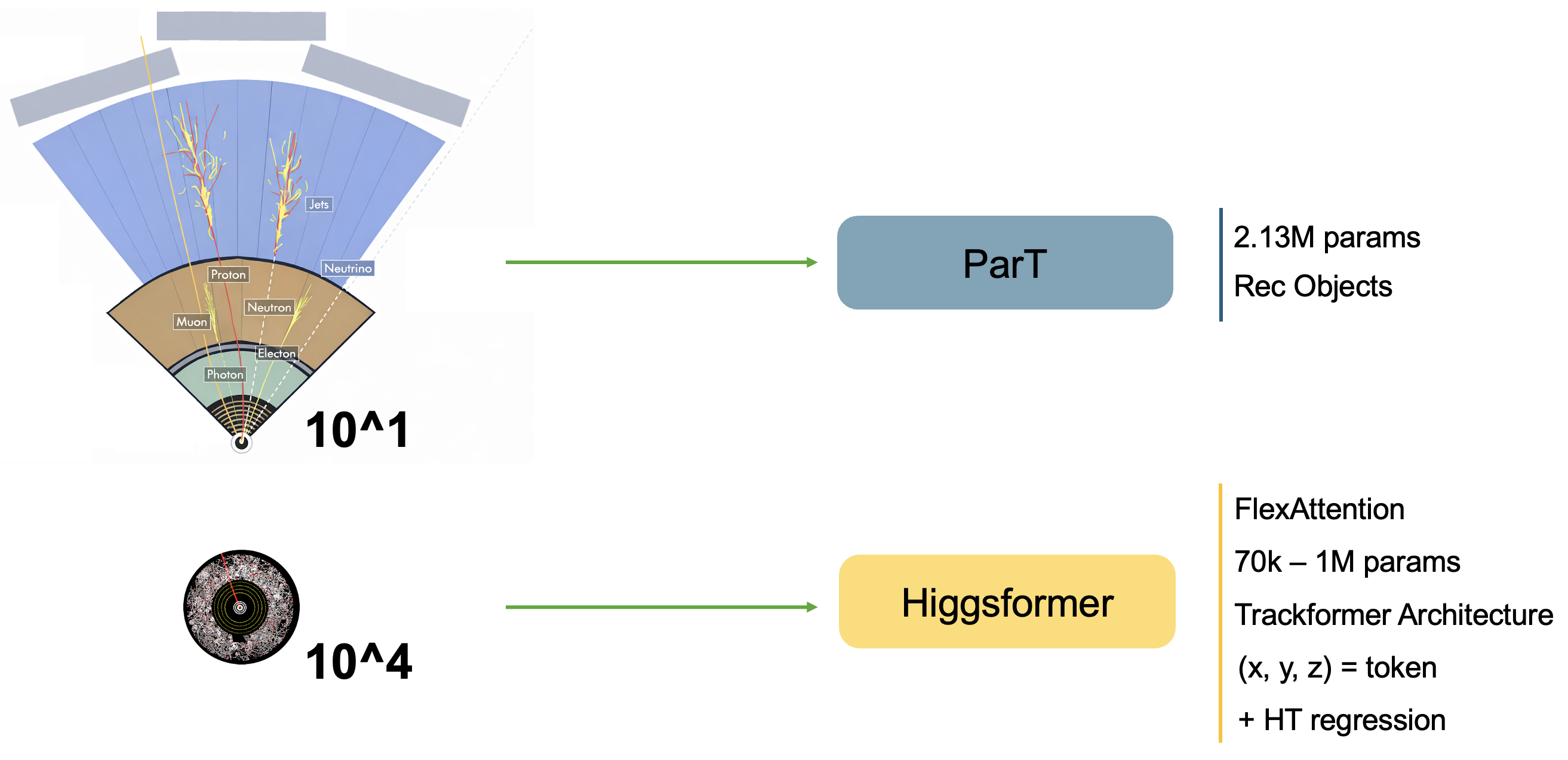}
    \caption{ParT (up) trained on reconstructed objects ($10^1$ inputs per event), and Higgsformer (bottom) trained on detector hits ($10^4$ inputs per event).}
    \label{fig:models}
\end{figure}

We evaluate two different models: Higgsformer and ParT, trained on and operating with raw detector level data and reconstructed object data, respectively, as shown in~\Cref{fig:models}.

\paragraph{Higgsformer} We introduce Higgsformer, a lightweight set-based Transformer operating directly on raw hit coordinates $(x,y,z)$. Each hit is treated as a token, embedded into a $d$-dimensional space, and processed with $L$ Transformer encoder layers. We present two models: Higgsformer-small and Higgsformer-big.

Higgsformer-small uses $L=2$ encoder layers, $H=4$ attention heads, and hidden size $d=32$. Input features are projected with a linear layer followed by LayerNorm. The encoded hit representations are aggregated with masked mean pooling to obtain an event-level representation, which is passed to a linear classifier for binary signal-background discrimination. For efficiency, Higgsformer-small uses FlashAttention~\cite{dao2022flashattention} and mixed precision training (AMP).

Higgsformer-big uses the EncCla Trackformer encoder backbone~\cite{Caron:2024cyo} adapted to multi-task learning. It employs $L=8$ layers, $H=8$ attention heads, and hidden size $d=128$. As in the Higgsformer-small model, hit features are first linearly embedded and then processed with Transformer encoder layers. For the classification branch, several aggregation strategies were studied, namely mean pooling, max pooling, and a learnable CLS token. No significant performance differences were observed between the global pooling strategies, and max pooling was used as the default choice. In addition to the binary classification head, Higgsformer-big includes an auxiliary $H_T$ regression head implemented as a two-layer MLP on the max-pooled event representation. This auxiliary objective is used to encourage the model to learn more physically meaningful event-level features. For memory-efficient attention on long sequences, Higgsformer-big uses FlexAttention~\cite{dong2024flex} with dynamic padding masks, and can be initialised from pretrained Trackformer encoder weights in the future.


Training uses binary cross-entropy for classification with the AdamW optimiser.  Higgsformer-small is trained with a learning rate of $10^{-4}$ for $30$ epochs with batch size $128$, ReduceLROnPlateau and early stopping with patience of $10$ epochs. Higgsformer-big is trained with binary cross-entropy and Huber loss for $H_T$ regression with equal weight contributions in the loss. Higgsformer-big is trained with learning rate $5*10^{-5}$, the other settings are unchanged. Dropout $(0.2)$ and weight decay are applied for regularisation for both models.

\paragraph{Particle Transformer (ParT)}
To benchmark against a state-of-the-art object-level model, we employ the Particle Transformer~\cite{Qu:2022mxj} for the $t\bar{t}H$ vs.\ $t\bar{t}$ classification task. The model operates on a sequence of reconstructed objects, using self-attention layers to capture correlations between particles. 
The detailed network configuration and hyperparameters follow our previous work~\cite{Builtjes:2022usj}. 
In Ref.~\cite{Builtjes:2022usj}, several physics-informed ParT variants were developed in which pairwise kinematic features and Standard Model interaction matrices were incorporated into the attention mechanism. In the present work, these extensions are intentionally omitted and the model is used in its vanilla self-attention form, in order to provide a clean object-level baseline for comparison with the hit-level Higgsformer approach.

    
    

\section{Experiments and Results}
\label{sec:experiments}

\paragraph{Performance at Dataset Sizes (Pileup 0) and Model Scale}

\begin{figure}[h]
    \centering
    \includegraphics[width=0.75\linewidth]{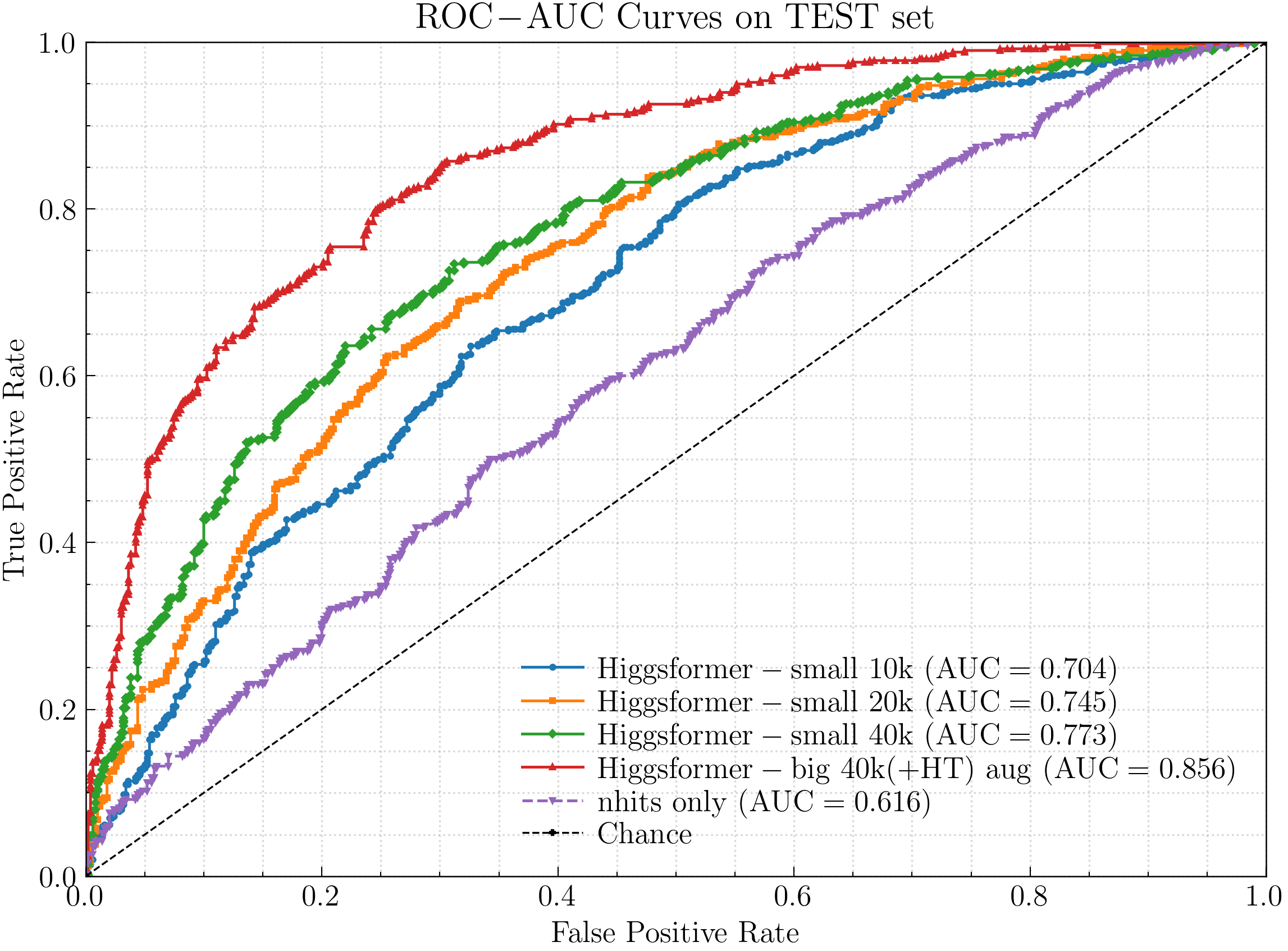}
    \caption{ROC curves and corresponding AUC values evaluated on the test set for Higgsformer-small models trained on datasets with increasing size ($10$k, $20$k, $40$k), and for Higgsformer-big trained on the $40$k dataset with augmentation.}
    \label{fig:dataset_size}
\end{figure}

We trained Higgsformer-small on PU\,0 datasets of 10k, 20k, and 40k events, using identical validation and test splits of $1$k events each. This corresponds to training set sizes of $8$k, $18$k, and $38$k events, respectively. 
We evaluate Higgsformer against a baseline classifier that uses only the number of hits per event, $n_{\text{hits}}$, as a separating feature.
The test ROC AUC increases monotonically with training size for Higgsformer (\Cref{fig:dataset_size}), consistent with scaling trends observed in Trackformers~\cite{Caron:2024cyo}.
Higgsformer-big yields a substantial further improvement over Higgsformer-small, demonstrating the benefit of model scale.

\paragraph{Performance at Pileup Level (Pileup 0, 5, and 20)}

\begin{figure}[h]
\centering
\includegraphics[scale=0.25]{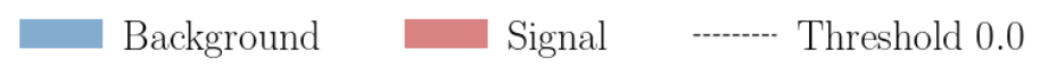}
\\
\begin{subfigure}{0.4\textwidth}
  \includegraphics[width=\linewidth]{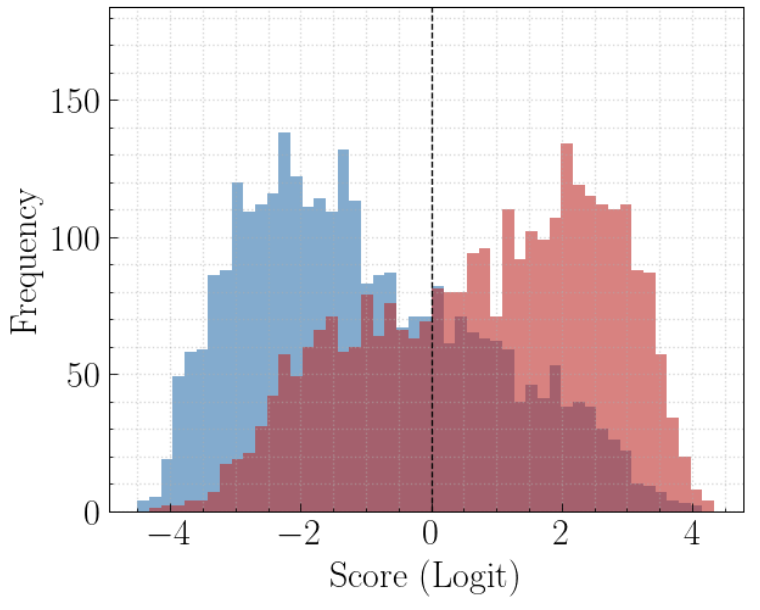}
  \caption{PU\,0 Higgsformer-small (AUC $=0.773$)}
  \label{fig:pileup_results_logit_p0}
\end{subfigure} \hfil
\begin{subfigure}{0.4\textwidth}
  \includegraphics[width=\linewidth]{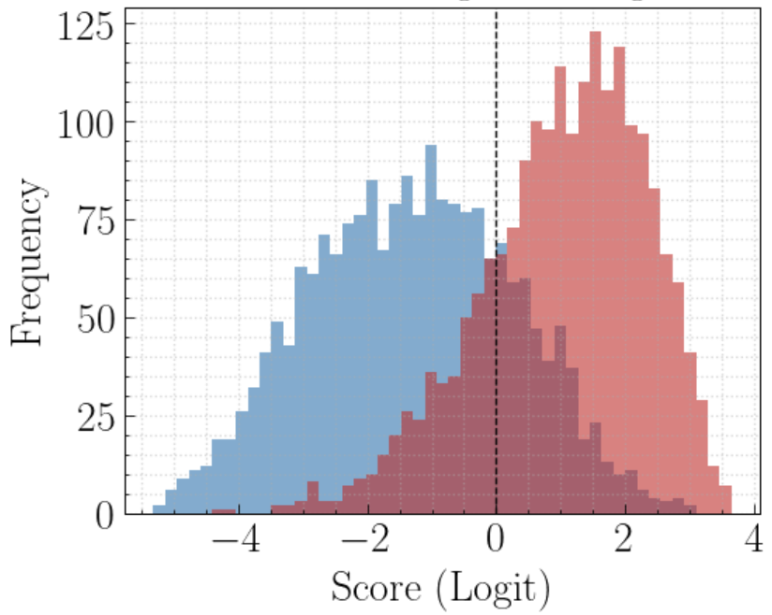}
  \caption{PU\,0 Higgsformer-big (AUC $=0.855$)}
  \label{fig:pileup_results_logit_p0_big}
\end{subfigure}\hfill
\\
\begin{subfigure}{0.4\textwidth}
  \includegraphics[width=\linewidth]{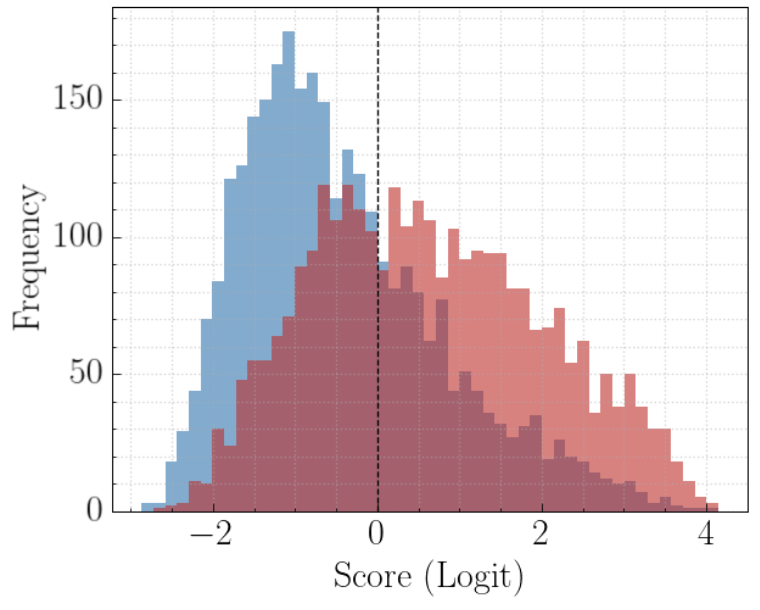}
  \caption{PU\,5 Higgsformer-small (AUC $=0.731$)}
  \label{fig:pileup_results_logit_p5}
\end{subfigure}\hfil
\begin{subfigure}{0.4\textwidth}
  \includegraphics[width=\linewidth]{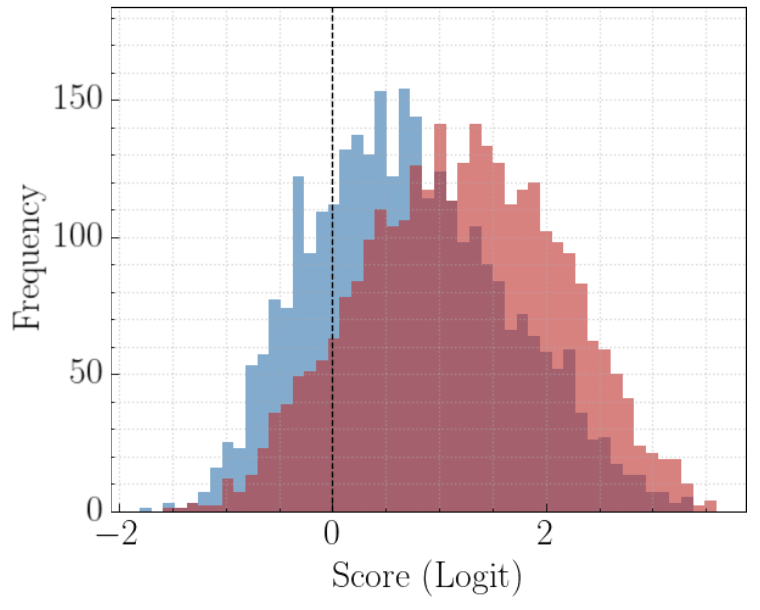}
  \caption{PU\,20 Higgsformer-small (AUC $=0.654$)}
  \label{fig:pileup_results_logit_p20}
\end{subfigure}
\caption{Classification output (Logit) histograms for pileup $0$ for Higgsformer-small (a) and Higgsformer-big (b) models, and pileup $5$ (c) and $20$ (d) for Higgsformer-small. Signal $t \overline{t} H$ events are shown in red and background $t\overline{t}$ events are shown in blue.}
\label{fig:pileup_results_logit} 
\end{figure}

\begin{figure}[h]
\centering
\begin{subfigure}{0.32\textwidth}
  \includegraphics[width=\linewidth]{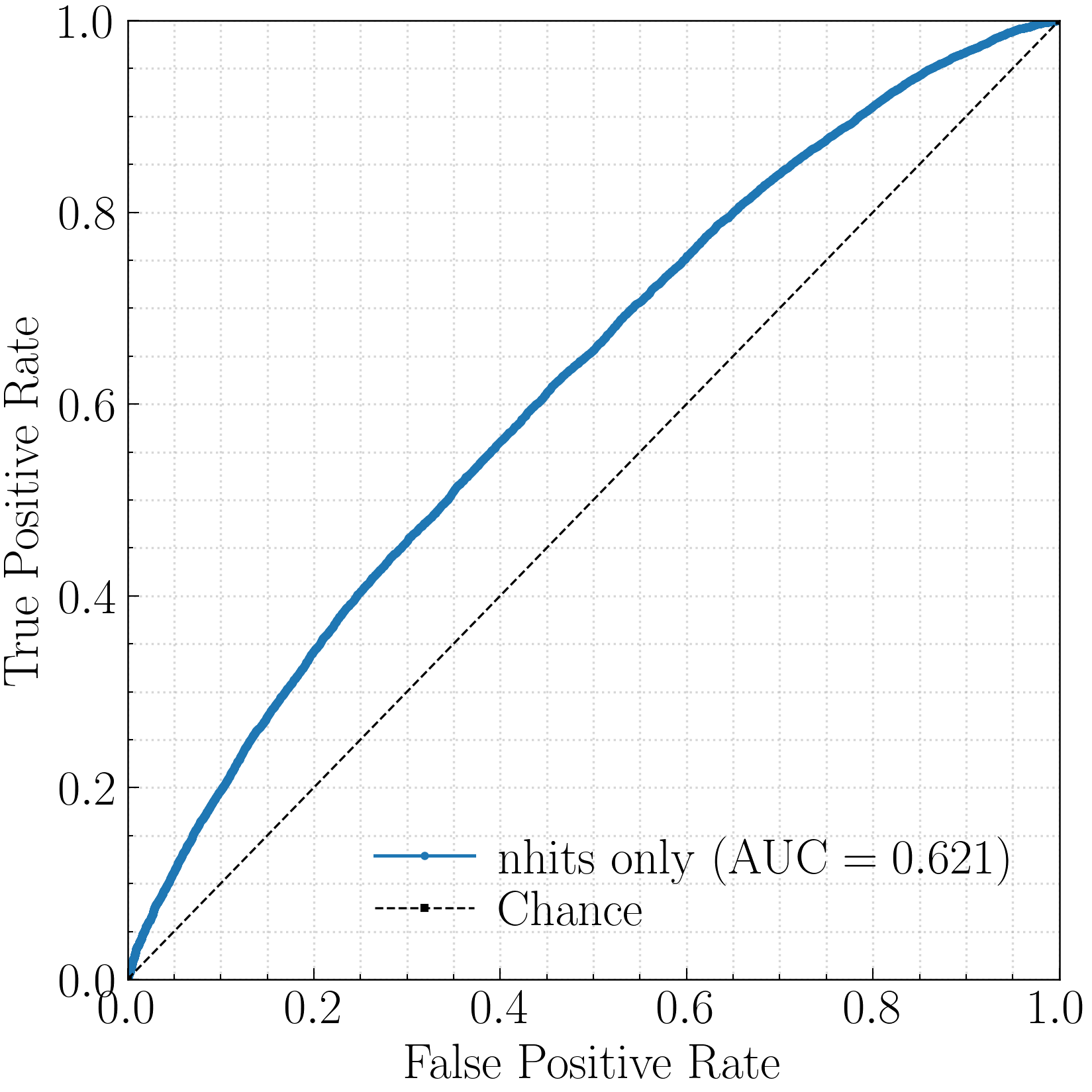}
  \caption{PU\,0 (AUC $=0.621$)}
  \label{fig:roc_nhits_pu0}
\end{subfigure}\hfill
\begin{subfigure}{0.32\textwidth}
  \includegraphics[width=\linewidth]{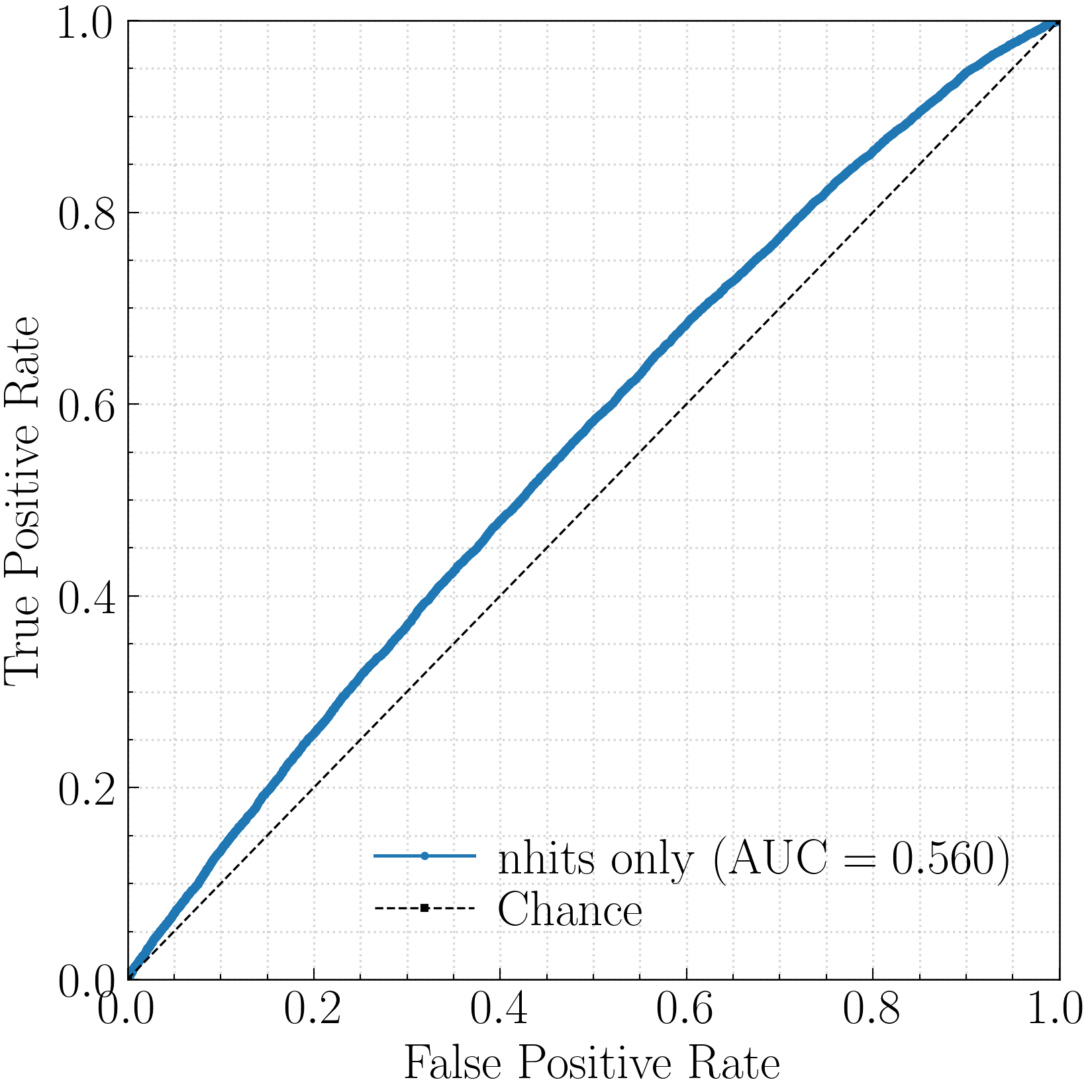}
  \caption{PU\,5 (AUC $=0.560$)}
  \label{fig:roc_nhits_pu5}
\end{subfigure}\hfill
\begin{subfigure}{0.32\textwidth}
  \includegraphics[width=\linewidth]{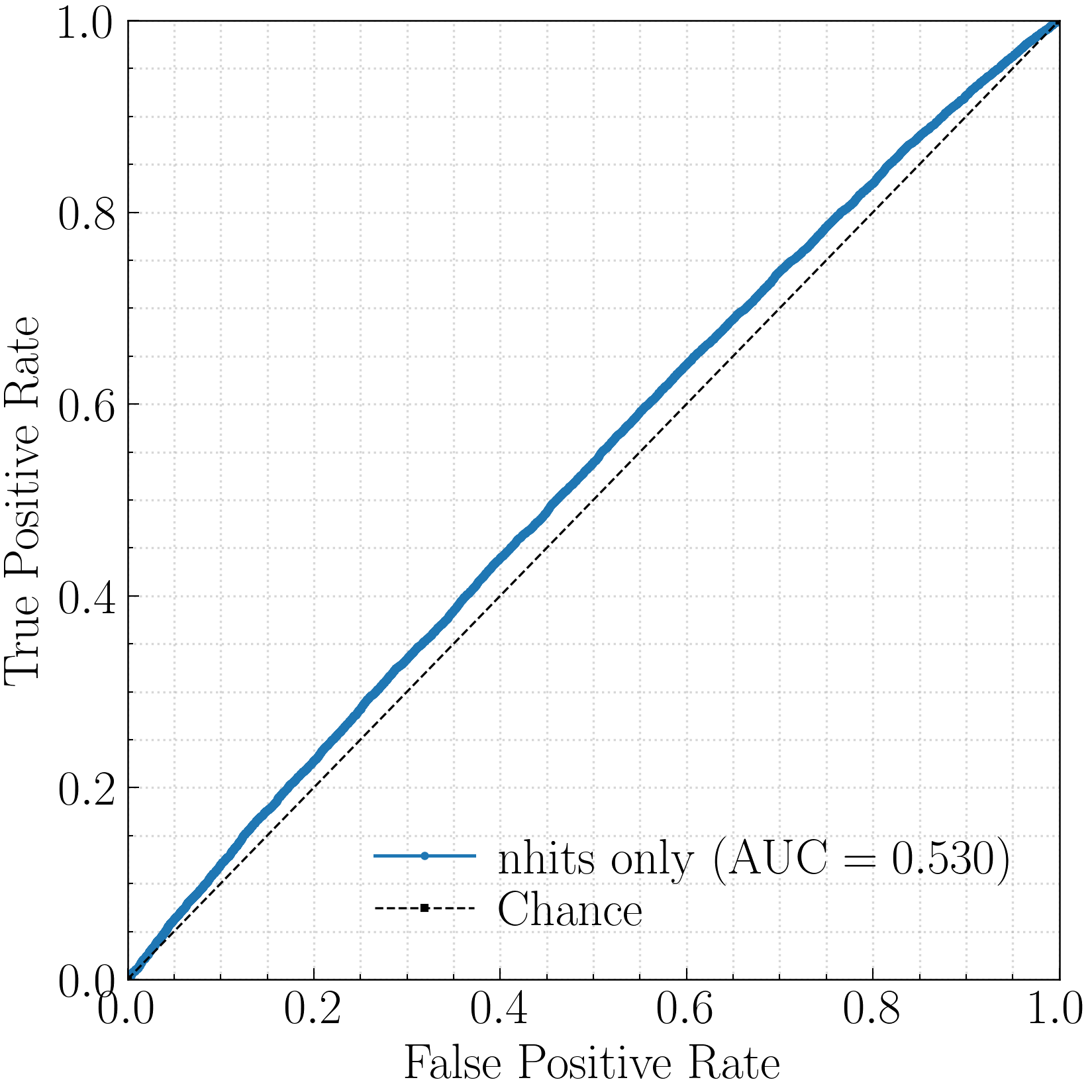}
  \caption{PU\,20 (AUC $=0.530$)}
  \label{fig:roc_nhits_pu20}
\end{subfigure}
\caption{ROC curves for a counts-only classifier using $n_{\text{hits}}$ at different pileup levels for all $40$k events.}
\label{fig:roc_nhits_all_pu}
\end{figure}

We trained three different Higgsformer-small models on the $38$k dataset generated with the varying pileup of $0$, $5$, and $20$, and evaluated those models on the 2k test set of respective pileup level. Performance degrades with pileup yet remains well above random. The logit histograms (Fig.~\ref{fig:pileup_results_logit}) show that the model still learns to separate classes even at PU\,20, while an $n_{\text{hits}}$-only baseline weakens dramatically with increasing pileup (Fig.~\ref{fig:roc_nhits_all_pu}). Additionally, we show the logit histogram for Higgsformer-big  trained on $38$k pileup $0$ dataset to show that it learns better class separation.

\paragraph{Geometric Data Augmentation}

\begin{table}[h]
\centering
\caption{Effect of geometric data augmentation on the test ROC AUC of Higgsformer-small at zero pileup for different  training set sizes. Values are reported as the mean over three independent training runs, with the standard deviation shown as the quoted uncertainty. The column `+ Aug' reports the AUC obtained with augmentation.}
\label{tab:aug_pu0}
\begin{tabular}{ccc}
\toprule
\textbf{Dataset size} & \textbf{Higgsformer-small} & \textbf{+ Aug} \\
\midrule
10k & 0.703(10) & \textbf{0.723(10)} \\
20k & 0.739(6) & \textbf{0.761(2)} \\
40k & 0.774(8) & \textbf{0.791(1)} \\
\bottomrule
\end{tabular}
\end{table}
We exploited detector symmetries via online $(x,y)$ $\phi$-rotations and a $z\!\to\!-z$ flip (train split only, val/test unchanged). These augmentations lead to consistent gains, aligning the training distribution with the known cylindrical and forward-backward symmetries, as shown in~\Cref{tab:aug_pu0}.

\paragraph{Comparison to Reconstruction Object Baselines}
\label{sec:obj-vs-hit}



\begin{figure}[]
  \centering
  \includegraphics[width=0.75\linewidth]{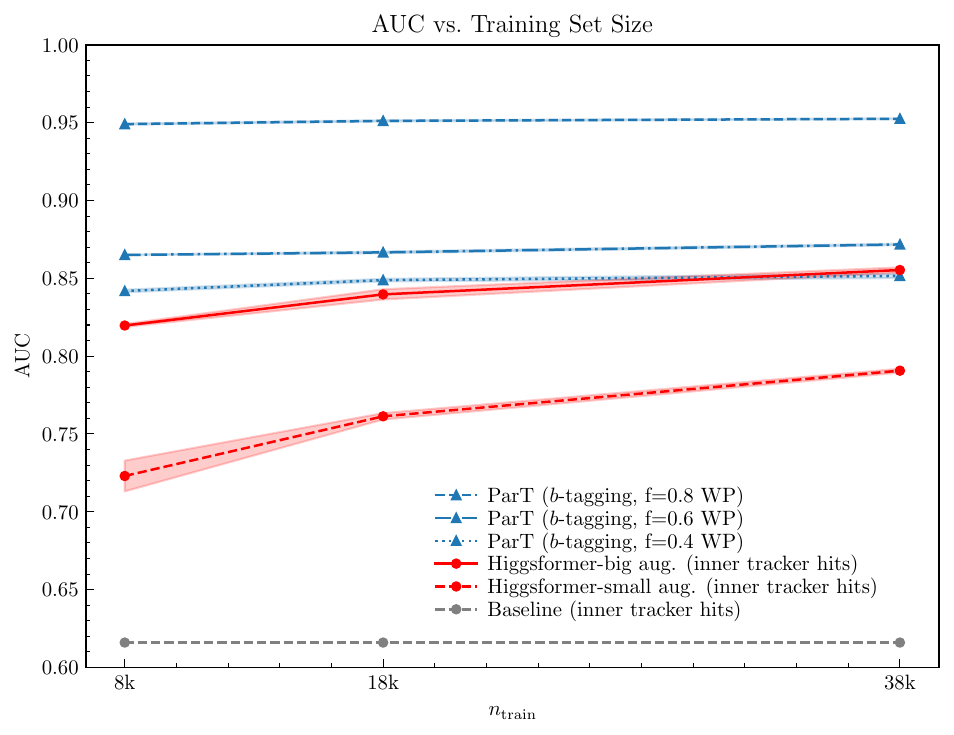}
  \caption{Test AUC as a function of training set size for ParT trained on reconstructed objects at three $b$-tagging WPs and for the augmented Higgsformer models trained on raw inner tracker hits. 
  Points show the mean AUC over three independent training runs, and the shaded bands indicate the standard deviation across these runs. 
  While the ParT performance is close to saturation, the augmented Higgsformer models continue to improve with scale. At $38$k training events, Higgsformer-big reaches an AUC comparable to ParT at the $f=0.4$ $b$-tagging WP. The horizontal dashed line shows the counts-only baseline ($\mathrm{AUC}=0.616$).}
  \label{fig:comparison_models}
\end{figure}

We compared our hit-level models against several classifiers trained on reconstructed objects using only inner tracker information. 
The ParT models operate on sequences of reconstructed physics objects derived from \texttt{Delphes}, and were evaluated at three $b$-tagging working points (WPs), corresponding to efficiencies of $0.4$, $0.6$, and $0.8$.
As shown in~\Cref{tab:obj_vs_hits} and~\Cref{fig:comparison_models}, all object-based models achieve high AUC values and exhibit signs of saturation with increasing training data. In contrast, Higgsformer-small and Higgsformer-big continue to improve with scale. Remarkably, Higgsformer-big reaches an AUC comparable to the object-based model at the $f=0.4$ working point.

\begin{table}[H]
\centering
\small
\caption{Test AUC as a function of training set size for the object-based model (ParT) and the hit-based models (Higgsformer-small and Higgsformer-big). ParT is trained on reconstructed physics objects, while Higgsformer operates directly on raw detector hits. The quoted uncertainties are the standard deviations computed from three independent training runs.}
\label{tab:obj_vs_hits}
\vspace{0.3em}
    \begin{tabular}{@{} ll ccc @{}}
      \toprule
      \multirow{2}{*}{\textbf{Model}} & \multirow{2}{*}{\textbf{Input}} & \multicolumn{3}{c}{\textbf{AUC @ $n_{\text{train}}$ (k)}} \\
      \cmidrule(lr){3-5}
       &  & \textbf{8} & \textbf{18} & \textbf{38} \\
      \midrule
    ParT$^a$ (\textit{WP = 80\%}) & Reco objects & 0.949(0) & 0.951(0) & 0.952(0) \\
    ParT$^a$ (\textit{WP = 60\%}) & Reco objects & 0.865(0) & 0.867(1) & 0.872(1) \\
    ParT$^a$ (\textit{WP = 40\%}) & Reco objects & 0.842(1) & 0.849(1) & 0.852(1) \\    
    \midrule
    Higgsformer-big$^b$ (\textit{augmented, only inner tracker})& \multirow{2}{*}{Raw detector hits} & 0.820(1) & 0.840(3) & 0.855(2) \\
    Higgsformer-small (\textit{augmented, only inner tracker})  &  & 0.723(10) & 0.761(2) & 0.791(1) \\   
    \bottomrule
\end{tabular}

\vspace{0.5em}

\begin{minipage}{0.92\textwidth}
  \footnotesize
  $^a$ Includes $\slashed{E}_T$, $\phi_{\slashed{E}_T}$, all jets and all $b$-jets. \\
  $^b$ Includes $H_T$ regression \\
\end{minipage}
\end{table}

\paragraph{Learned Feature Exploration}
\label{sec:learned_features}
A counts-only classifier based on $n_{\text{hits}}$ provides limited discrimination that diminishes with pileup, while Higgsformer maintains a substantial advantage, as shown in~\Cref{fig:pileup_results_logit,fig:roc_nhits_all_pu}.


Crucially, we verify that the model learns Higgs-related structure on $98$ unseen events. Using a leave-one-hit-out importance measure and a HepMC\,\textrightarrow\,ROOT\,\textrightarrow\,CSV matching chain to tag hits from final Higgs descendants, we find that Higgs-tagged hits carry systematically higher importance than non-Higgs hits, with the effect strengthening at larger training size, as shown in~\Cref{tab:higgs_importance}.

\begin{table}[H]
\centering
\caption{Higgs and non-Higgs hits importance with \textit{Higgsformer-small} $38$k vs.\ $8$k augmented data.}
\label{tab:higgs_importance}
\begin{tabular}{lrr}
\toprule
\textbf{Metric} & \textbf{38k augmented} & \textbf{8k augmented} \\
\midrule
Mean avg(H)$^a$ & 0.005777 & 0.002217 \\
Mean avg(non-H)$^b$ & 0.004560 & 0.001965 \\
Mean $\Delta$avg (H$-$non-H)$^c$ & 0.001218 & 0.000252 \\
Events with avg(H) $>$ avg(non-H) & 72/98 & 51/98 \\
\bottomrule
\end{tabular}
\vspace{0.5em}

\begin{minipage}{0.92\textwidth}
  \footnotesize
  $^a$ avg(H) – average importance of Higgs-related hits in an event. \\
  $^b$ avg(non-H) – average importance of hits in an event, originating from other particles. \\
  $^c$ $\Delta$avg (H$-$non-H) – difference between the average importance of Higgs-related and non-Higgs-related hits in an event.
\end{minipage}
\end{table}

\begin{figure}[H]
\centering
\begin{subfigure}{0.49\textwidth}
  \includegraphics[width=\linewidth]{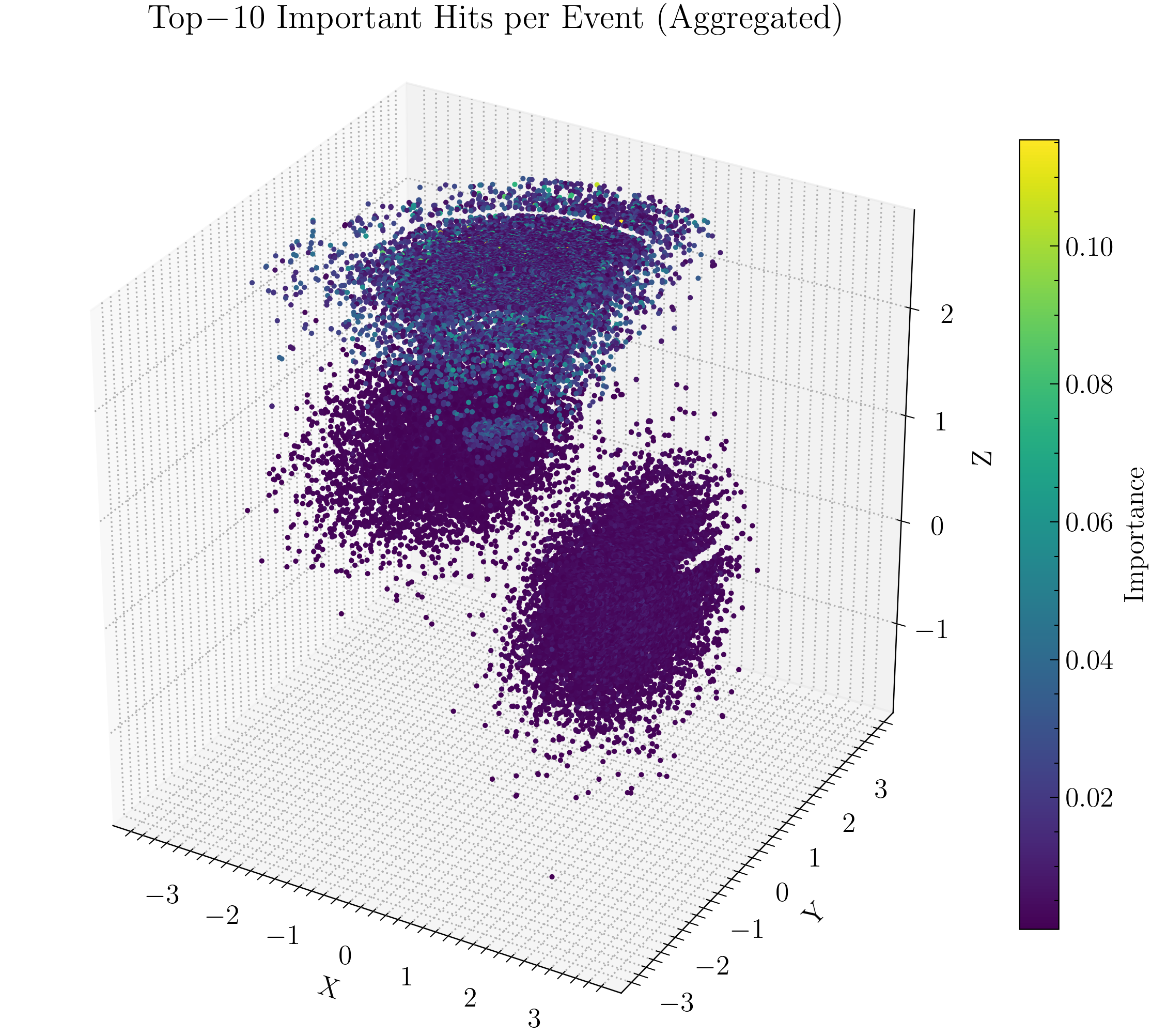}
  \caption{Top-$10$ important hits for Higgsformer-small 10k.}
  \label{fig:topn_3d_10k}
\end{subfigure}\hfill
\begin{subfigure}{0.49\textwidth}
  \includegraphics[width=\linewidth]{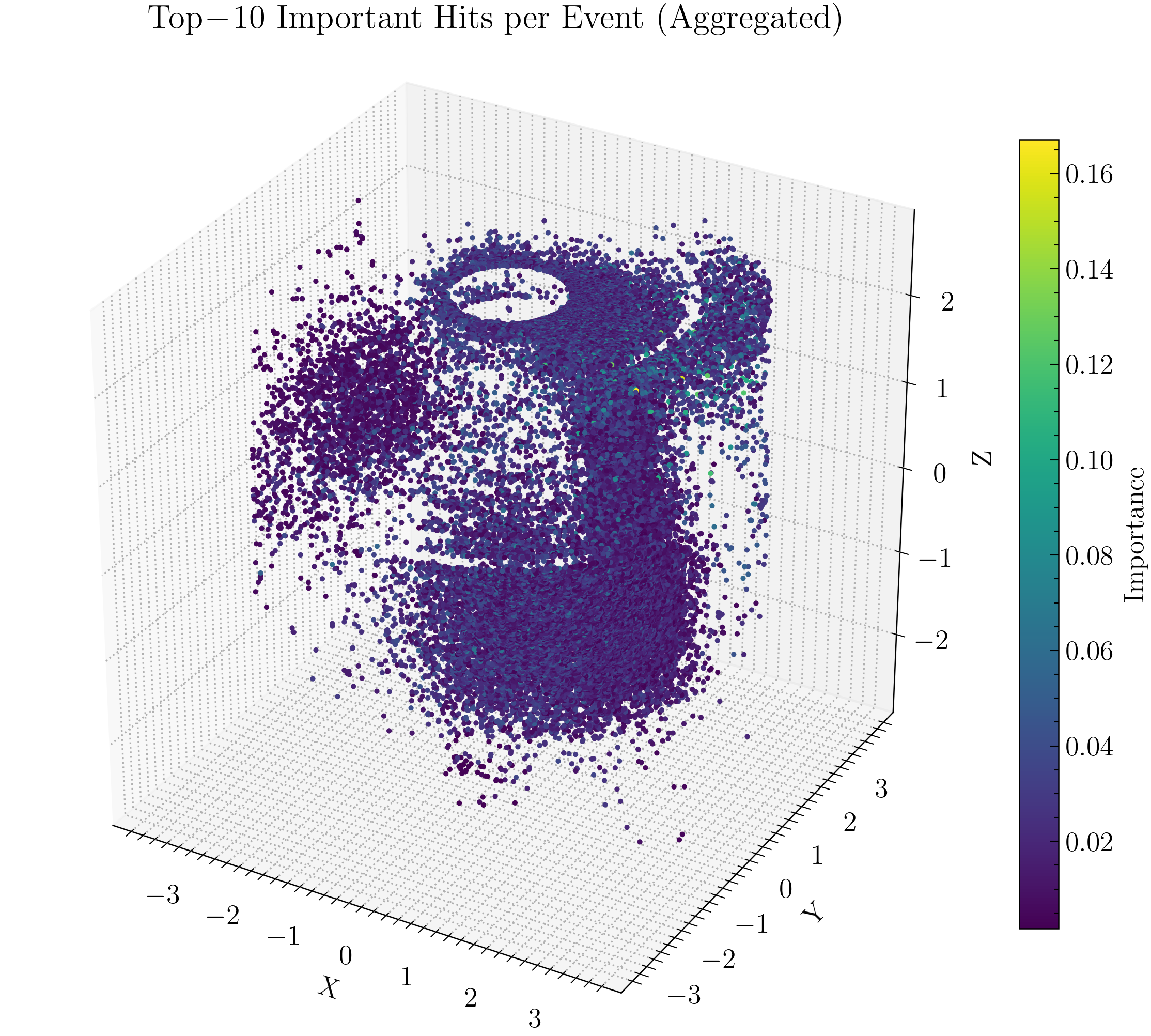}
  \caption{Top-$10$ important hits for Higgsformer-small 40k.}
  \label{fig:topn_3d_40k}
\end{subfigure}\hfill
\caption{Top-$10$ important hits (3D) for Higgsformer-small trained with $10$k (left) and $40$k (right) training data.}
\label{fig:topn_3d}
\end{figure}

The 3D distributions of the ten most important hits per event (Fig.~\ref{fig:topn_3d}), ranked by their contribution to the final loss, were analyzed to examine how the model captures the symmetries of the system. A rotational symmetry in the $x$-$y$ plane is expected. As shown in~\Cref{fig:topn_3d}, Higgsformer progressively learns these detector symmetries as the size of the training dataset increases.

Overall, these results indicate that Higgsformer exploits physically meaningful geometric information and increasingly focuses on hits originating from the Higgs decay as the amount of training data increases.


\paragraph{Time Complexity}
Finally, we evaluated the runtime performance of the models. While classical CPU-based tracking typically requires about 1~s per event, the inference times of Higgsformer-small and Higgsformer-big are below $2~ms$ and $10~ms$, respectively, on an NVIDIA A100 GPU for pileup\,=\,$0$. This represents a speed-up of several orders of magnitude.

\section{Conclusions}
\label{sec:conclusions}
Our results demonstrate that the Higgsformer model, trained solely on inner tracker hits, can effectively separate $t\bar{t}H$ from $t\bar{t}$ events, achieving a test ROC AUC of $0.855$ at zero pileup for our best model. Performance is likely limited by the size of the training dataset and the degree of model optimisation.


The analysis of learned features in Section~\ref{sec:learned_features} reveals that the network does not rely on simple proxies such as the total number of hits. Instead, it systematically assigns higher importance to hits traceable to Higgs decay products. With increasing training data, high-importance hits form more symmetric and cylindrical patterns, suggesting that Higgsformer learns both detector geometry and Higgs-specific structures.


While object-level models still outperform Higgsformer for $b$-tagging efficiencies $f>0.4$, our central conclusion is that non-trivial event-level discrimination can be learned directly from raw inner tracker hits under detector-consistent conditions. These results demonstrate for the first time that modern attention-based architectures can extract meaningful discriminative information directly from hit-level data. This represents a promising step toward hit-level classification in high energy physics, i.e.\ approaches that do not require explicit intermediate reconstructed objects as inputs.

In addition to improving the ML model architecture, future work will focus on scaling to larger datasets, integrating additional subdetectors, and testing the approach under more realistic detector conditions and high-pileup environments.

\acknowledgments
We thank Dr. Paul Gessinger for helpful contributions on ACTS and Fatras. R. RdA is supported by PID2020-113644GB-I00 from the Spanish Ministerio de Ciencia e Innovación and by PROMETEO/2022/69 from the Spanish GVA.



\clearpage
\bibliographystyle{JHEP}
\bibliography{refs.bib}

\end{document}